\documentclass[journal]{IEEEtran}
\usepackage{cite}

\usepackage{subfigure}
\ifCLASSINFOpdf
\usepackage[pdftex]{graphicx}
  \graphicspath{{../pdf/}{../jpeg/}}
  \DeclareGraphicsExtensions{.pdf,.jpeg,.png}
\else
  \usepackage[dvips]{graphicx}
  \graphicspath{{../eps/}}
  \DeclareGraphicsExtensions{.eps}
\fi
\usepackage{amsmath}
\usepackage{cases}
\usepackage{amsfonts}
\usepackage{subeqnarray}
\usepackage{longtable}
\usepackage{supertabular}
\usepackage{setspace}
\usepackage[ruled,linesnumbered]{algorithm2e}
\makeatletter
\newcommand{\nosemic}{\renewcommand{\@endalgocfline}{\relax}}
\newcommand{\dosemic}{\renewcommand{\@endalgocfline}{\algocf@endline}}
\let\oldnl\nl
\newcommand{\nonl}{\renewcommand{\nl}{\let\nl\oldnl}}
\makeatother
\usepackage[export]{adjustbox} 
\usepackage{booktabs}
\usepackage{setspace}
\usepackage{xcolor}
\usepackage{mathrsfs}
\usepackage{amsmath}
\usepackage{array}
\usepackage{amssymb}
\usepackage{amsthm}
\usepackage{microtype}
\usepackage{url}
\usepackage{amsfonts,amssymb}
\usepackage{bbm}
\usepackage{mathtools}

\allowdisplaybreaks
\begin{document}
\setstretch{0.92}
\title{\textls[-25]{Iterative Decomposition of Joint Chance Constraints in OPF}}
\author{Mengshuo~Jia,\!~\IEEEmembership{Student~Member,~IEEE,} 
Gabriela~Hug,~\IEEEmembership{Senior~Member,~IEEE,} and
Chen~Shen,~\IEEEmembership{Senior~Member,~IEEE} 

}
        

\maketitle

\begin{abstract}
In chance-constrained OPF models, joint chance constraints (JCCs) offer a stronger guarantee on security compared to single chance constraints (SCCs). Using Boole's inequality or its improved versions to decompose JCCs into SCCs is popular, yet the conservativeness introduced is still significant. In this letter, a non-parametric iterative framework is proposed to achieve the decomposition of JCCs with negligible conservativeness. An adaptive risk allocation strategy is also proposed and embedded in the framework. Results on an IEEE test case show that the conservativeness using the framework is nearly eliminated, thereby reducing the generation cost considerably. 
\end{abstract}
\IEEEpeerreviewmaketitle

\vspace{-0.3cm}
\section{Introduction}

\IEEEPARstart{A}{}wide variety of chance-constrained OPF (CC-OPF) models that account for single chance constraints (SCCs) have been developed \cite{7862254}, yet SCCs ignore the simultaneous violation situations in the system \cite{BAKER2017230}. Instead, joint chance constraints (JCCs) offer a stronger guarantee on the overall system security \cite{9122389}, as all of the constraints are considered concurrently \cite{7224103}.

Dealing with JCCs is generally more challenging than SCCs, because the JCC is composed of both the marginal and joint violating probabilities of SCCs. Adopting Boole's inequality to separate the JCC into SCCs is a popular approach \cite{HongJeff}. Nevertheless, Boole's inequality brings obvious conservativeness no matter whether the uniform \cite{GROSSO2014504} or optimal \cite{4739221} risk allocation is used for each SCC, as the joint violating probabilities of SCCs is neglected in both cases. To reduce conservativeness, an improved version of Boole's inequality has been derived through estimating the joint violating probability of all SCCs in \cite{BAKER2017230}. Hereafter, an improving bound method is further developed to approximate the equivalent decomposition of JCCs \cite{8662704}. Briefly, this method first identifies binding constraints, then estimates the joint violating probability for any desired combination of binding SCCs and finally decomposes the JCC into binding SCCs using the probability estimation. Consequently, the OPF solution is obtained. This method does bring a fresh perspective regarding transforming the JCC into SCCs. However, it may lack thorough considerations in two aspects. Namely, 1) this method ignores the interdependence among the constraint classification, probability estimation, and OPF solution, meaning that the solution may be incompatible with the classification and estimation, thus failing to satisfy the original JCC; 2) it adopts uniform risk levels for all SCCs, which leads to obvious conservative results. 

This letter also focuses on solving the joint CC-OPF through decomposing JCCs into SCCs. The key contribution is proposing a non-parametric iterative framework to realize the decomposition with negligible conservativeness, thereby achieving a less costly solution. The implementation of the framework is simple and straightforward as it does not need tuning parameters or using additional algorithms to search for appropriate parameters. Compared to the related state-of-the-art methods \cite{BAKER2017230, 8662704}, the innovations of the framework are twofold. First, an iterative structure is developed to gradually reach a fixed point where the constraint classification, probability estimation, and OPF solution match each other. In the end, both the classification and estimation are accurate given the current, stable optimal solution. Second, an adaptive risk allocation strategy is proposed for further relaxation. Accordingly, the overall conservativeness is almost removed.

In the following, Section II proposes the non-parametric iterative framework. Section III performs case studies. Section IV concludes this letter and gives an outlook on future work.
\vspace{-0.3cm}
\section{Proposed Methodology}
\subsection{Joint CC-OPF}
The joint CC-OPF considered in this letter aims to minimize the generation cost: 
\begin{align}
	\min_{\boldsymbol{g}_{t} } \quad f = \sum\nolimits_{t \in \mathcal{T} } C (\boldsymbol{g}_t) \label{P1}
\end{align}
where $\boldsymbol{g}_t $ is a vector consisting of active power outputs from $N_g$ dispatchable generators at time $t$. $\mathcal{T}$ is the set of time instances. The typical quadratic cost function $C(\cdot)$ \cite{7862254} is adopted here.
 
At time $t$, the capacity constraints of the generators are:
\begin{align}
	\boldsymbol{g}^- \leq  \boldsymbol{g}_{t} \leq \boldsymbol{g}^+  
\end{align}
while the ramp rate constraints are:
\begin{align}
	\boldsymbol{r}^- \leq \boldsymbol{g}_{t+1} - \boldsymbol{g}_{t}   \leq \boldsymbol{r}^+ 
\end{align}
where $\boldsymbol{g}^+$/$\boldsymbol{g}^-$ and $\boldsymbol{r}^+$/$\boldsymbol{r}^-$ are the upper/lower bounds of generators' capacities and ramp rates, respectively. 

Given the non-dispatchable wind power, the supply-demand constraint at time $t$ is formulated as a SCC \cite{8254387}: 
\begin{align}
	\mathbb{P}\{\mathbbm{1}_g^\top \boldsymbol{g}_{t} + \mathbbm{1}_w^\top \boldsymbol{w}_{t} - \mathbbm{1}_d^\top \boldsymbol{d}_{t} \geq 0 \} \geq 1 - \epsilon \label{supply-demand}
\end{align}
where vector $\boldsymbol{w}_{t}$ consists of $N_w$ uncertain active wind power injections, $\boldsymbol{d}_{t}$ is a vector including $N_d$ forecasted active loads, and $\mathbbm{1}$ denotes the all-ones vector with the appropriate size. Vector $\boldsymbol{w}_{t}$ is further detailed by \eqref{w} using the forecasted value $\boldsymbol{\overline{w}}_t$ and the uncertain forecast error $\boldsymbol{\delta}_t$:
\begin{align}
	\boldsymbol{w}_{t} = \boldsymbol{\overline{w}}_t + \boldsymbol{\delta}_t \label{w}
\end{align}
Note that this letter ignores the load uncertainty as it is relatively small compared to the uncertainty of wind power \cite{7565532}. 

At time $t$, the capacity constraint of transmission lines belonging to set $\mathcal{L}$ is modeled as a JCC:
\begin{align}
	\mathbb{P}\{  l_{i}^- \leq l_{it} \leq l_{i}^+, \forall i \in \mathcal{L} \} \geq 1 - \alpha  \label{JCC}
\end{align}
which enforces the corresponding active line flows to be within their upper ($l_{i}^+$) and lower ($l_{i}^-$) limits simultaneously with a probability of at least $1 - \alpha$. Similar to \cite{8662704,7862254,BAKER2017230,7224103,9122389}, this letter also adopts a linear power flow model:
\begin{align}
	l_{it} = \boldsymbol{\lambda}_{ig}^\top \boldsymbol{g}_{t} + \boldsymbol{\lambda}_{iw}^\top \boldsymbol{w}_{t} +\boldsymbol{\lambda}_{id}^\top \boldsymbol{d}_{t} \quad \forall i \in \mathcal{L} \label{l}
\end{align}
where $\boldsymbol{\lambda}_{ig}$, $\boldsymbol{\lambda}_{iw}$, and $\boldsymbol{\lambda}_{id}$ are vectors composed of the corresponding DC power transfer distribution factors.

Once the JCC is decomposed into SCCs, the optimal solution of the consequent OPF is obtainable using the method in \cite{7862254}. 

\vspace{-0.3cm}
\subsection{JCC Decomposition} 
Let $|\cdot|$ denote the cardinality function. There are $2|\mathcal{L}|$ single constraints contained in \eqref{JCC} at time $t$. To simplify notation, $x_n$ and $x_n^+$ are introduced. That is, $\forall i \in \mathcal{L}$, if $n = 2i-1$ then $x_n = l_{it}$ and $ x_n^+ = l_i^+$; if $n = 2i$ then $x_n = -l_{it}$ and $ x_n^+ = -l_i^-$.
Accordingly, any violation event corresponding to a single constraint in \eqref{JCC} can be uniformly expressed by $y_n$:
\begin{align}
	y_n \coloneqq \left\{x_n > x_n^+\right\} \quad n \in \mathcal{N} \label{y}
\end{align}
where $\mathcal{N}=\{1,...,2|\mathcal{L}|\}$. In fact, only some of the events are possible, i.e., the corresponding single constraints are binding. Define the index set of possible events as $\mathcal{N}_{p}$. Then
\begin{align}
	\mathbb{P}\{  l_{i}^- \leq l_{it} \leq l_{i}^+, \forall i \in \mathcal{L} \} = 1 - \mathbb{P}\left\{ \bigcup\nolimits_{n \in \mathcal{N}_p }y_n \right\}   \label{JCC-event}
\end{align}
where 
\begin{align}
	\label{probability}
	\mathbb{P}\left\{ \bigcup\nolimits_{n \in \mathcal{N}_p }y_n \right\}& =   \sum\nolimits_{n \in \mathcal{N}_p}\mathbb{P}\left(y_n\right)\!  -\! \sum\nolimits_{\substack{n<m\\ n,m \in \mathcal{N}_p}}\mathbb{P}\left(y_n\bigcap y_m \right) \notag \\
	& \ \ \ \ 	 + \ \cdots \  (-1)^{|\mathcal{N}_p|-1}\mathbb{P}\left(\bigcap\nolimits_{n \in \mathcal{N}_p }y_n \right) \notag \\
	& = \sum\nolimits_{n \in \mathcal{N}_p}\mathbb{P}\left(y_n\right) - E
\end{align}
according to the inclusion-exclusive principle \cite{8662704}. Notice that $E$ is the aggregation of the joint violating probabilities for all of the combinations of possible events. Consequently, \eqref{JCC} is equivalent to 
\vspace{-0.1cm}
\begin{align}
	\sum\nolimits_{n \in \mathcal{N}_p}\mathbb{P}\left(y_n\right) \leq \alpha + E 
\end{align}
Once $\mathcal{N}_p$ is identified and $E$ is estimated, then together with risk allocation, the JCC can be decomposed into SCCs:
\begin{align}
	\mathbb{P}\left(y_n\right)\leq \beta_n (\alpha + E) \quad  \forall n \in \mathcal{N}_p \label{SCC}
\end{align}
where $\beta_n$ is the risk allocation factor and $\sum\nolimits_{n=1}^{|\mathcal{N}_p|}\beta_n=1$.

In the following, a framework with adaptive $\beta_n$ is proposed to realize the above decomposition. 

\vspace{-0.4cm}
\subsection{Proposed Framework}
First, it is worth noting that the true $\mathcal{N}_p$ and $E$ not only depend on the wind power uncertainty but also rely on the final generator outputs, i.e., the optimal solution of the joint CC-OPF. Ideally, the optimal solution as well as the wind power uncertainty should be leveraged to identify $\mathcal{N}_p$ and  estimate $E$. However, the optimal solution is not obtainable unless $\mathcal{N}_p$ and $E$ are known --- the typical \textit{Chicken or the Egg} causality dilemma.

To break the deadlock, a non-parametric iterative framework is proposed here whereas the related discussions are provided in the next subsection. The framework consists of 6 steps.

\subsubsection{Samples Generation}
For each time instance, substitute $N_s$ samples of the forecast error into \eqref{w} to generate wind power samples, each of which is denoted by $\boldsymbol{w}_t^s$, where $s=1,...,N_s$.

\subsubsection{Pre-solving}
The joint CC-OPF is pre-solved using Boole's inequality, i.e., the JCC is separated into SCCs by  
\begin{align}
	\mathbb{P}\left(y_n\right)\leq \alpha \left. \big/ \right. |\mathcal{N}| \quad  \forall n \in \mathcal{N} \label{SCC-Boole} 
\end{align}	
The corresponding optimal solution is denoted by $\boldsymbol{g}_t^*$ ($t \in \mathcal{T}$).

\subsubsection{Classification}
Substitute $\boldsymbol{g}_t^*$ and $\boldsymbol{w}_t^s$ into \eqref{l} to generate $N_s$ line flows and concurrently $N_s$ samples for $x_n$, each of which is represented by $x_n^s$. Then, the marginal violating probability of each single event is estimated by
\begin{align}
	\mathbb{P}\left(y_n\right) \approx \hat{\mathbb{P}}\left(y_n\right) = \sum\nolimits_{s=1}^{N_s}\Gamma\left(x_n^s-x_n^+\right)/N_s \quad \forall n \in \mathcal{N}  \label{Pm}
\end{align}
where $\Gamma(a) =1$ if $a>0$ and otherwise 0. The indices where $\hat{\mathbb{P}}\left(y_n\right)>0$ ($\forall n \in \mathcal{N} $) form $\mathcal{N}_{p} $.

\subsubsection{Estimation}
Define $\mathcal{N}_{p}^\star \subseteq \mathcal{N}_{p} $ as a combination of possible violation events. The corresponding $x_n^s$ ($n \in \mathcal{N}_{p}^\star $) form vector $\boldsymbol{x}^s$, while $x_n^+$ ($n \in \mathcal{N}_{p}^\star $) compose $\boldsymbol{x}^+$. The joint violating probability of events in $\mathcal{N}_{p}^\star$ is estimated by 
\begin{align}
	\mathbb{P}\left(\bigcap\nolimits_{n \in \mathcal{N}_{p}^\star}y_n\right) \approx \sum\nolimits_{s=1}^{N_s}\Gamma\left[\boldsymbol{x}^s-\boldsymbol{x}^+\right]/N_s \label{Pj}
\end{align}
Then, $E$ is obtained by aggregating the estimated joint violating probabilities of all the combinations of possible events.

\subsubsection{Adaptive Risk Allocation}
The adaptive risk allocation factor is designed as follows:
\begin{align}
	& \beta_n = \hat{\mathbb{P}}\left(y_n\right) \left. \big/ \right. \sum\nolimits_{m \in \mathcal{N}_p}\hat{\mathbb{P}}\left(y_m\right) \quad \forall n \in \mathcal{N}_{p}
\end{align}

\subsubsection{Re-solving}
The joint CC-OPF is re-solved using the decomposition in \eqref{SCC} with the latest  $\mathcal{N}_p$, $E$, and  $\beta_n$. Correspondingly, the previous solution $\boldsymbol{g}_t^*$ ($t \in \mathcal{T}$) is updated. Hereafter, steps 3 to 6 are repeated until the relative variation of the optimal objective function value is negligible.
\vspace{-0.3cm}
\subsection{Discussions on the Proposed Framework}
The proposed adaptive risk allocation strategy can satisfy SCCs' actual requirements for risk thresholds, thereby reducing conservativeness. Specifically, a larger $\hat{\mathbb{P}}\left(y_n\right)$ means that the line flow is more likely to violate its physical limits. Hence, the corresponding SCC automatically obtains a looser upper bound. On the contrary, the SCC whose line flow hardly exceeds the limits gets a tighter upper bound, yet not much effort is required to satisfy such a bound. As a comparison, an identical allocation of $\beta_n$ puts too much pressure on flows that are easily over-limit but offers unnecessary room for flows that often stay within limits, thus causing obvious conservativeness.

Second, the interdependence among $\mathcal{N}_p$, $E$, and $\boldsymbol{g}_t^*$ has been taken into account. That is, the current $\boldsymbol{g}_t^*$ is used to update $\mathcal{N}_p$ and $E$, while the updated $\mathcal{N}_p$ and $E$ are again applied to modify $\boldsymbol{g}_t^*$. In the end, both $\mathcal{N}_p$ and $E$ are accurate in terms of the stable optimal solution. Ignoring this interdependence leads to possible conflicts among the classification, estimation, and solution. Accordingly, the original JCC may not be satisfied given the solution.

Third, although samples are used for the classification and estimation, the corresponding computational burden is low, because for each sample, only the linear function in \eqref{l} is required to be computed, instead of calculating a whole OPF problem as done in \cite{8662704}.
 
\vspace{-0.3cm}
\section{Case Study}
A modified IEEE 39-bus system with $N_g=10$ and $N_w=8$ is adopted for testing. Set $\mathcal{L}$ is composed of the lines directly connected to the wind power injections. Besides, $|\mathcal{T}|=24$, $N_s=20000$, and $\epsilon$ in \eqref{supply-demand} is chosen as $10^{-4}$. Note that both high and low load levels with different $\alpha$ are used for experiments, which are indicated in the caption of the figure/table.
\begin{figure}[h]
\centering  
\subfigure[]{ 
\includegraphics[width=1.61in]{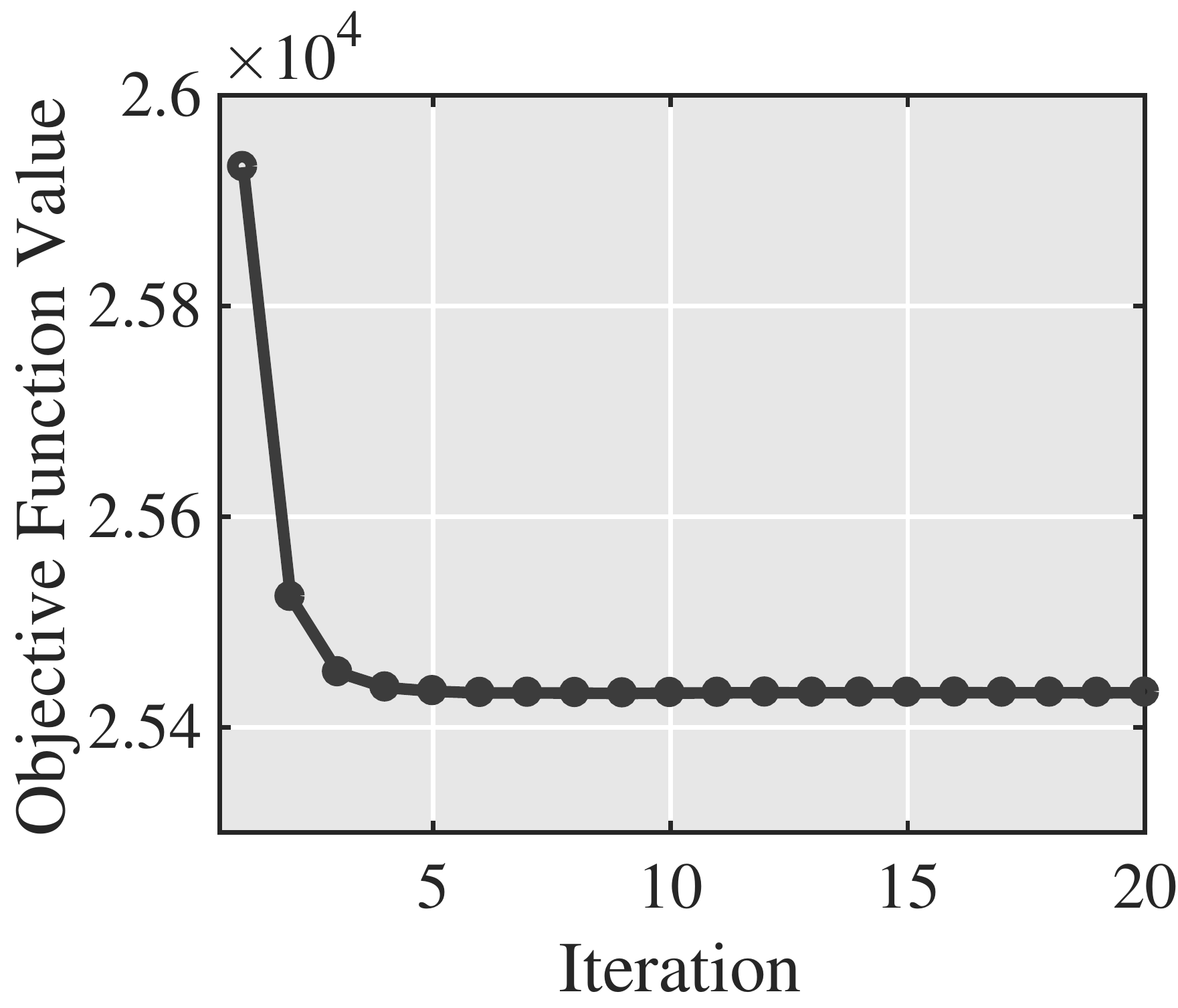} 
} 
\vspace{-0.12in}
\subfigure[]{ 
\includegraphics[width=1.61in]{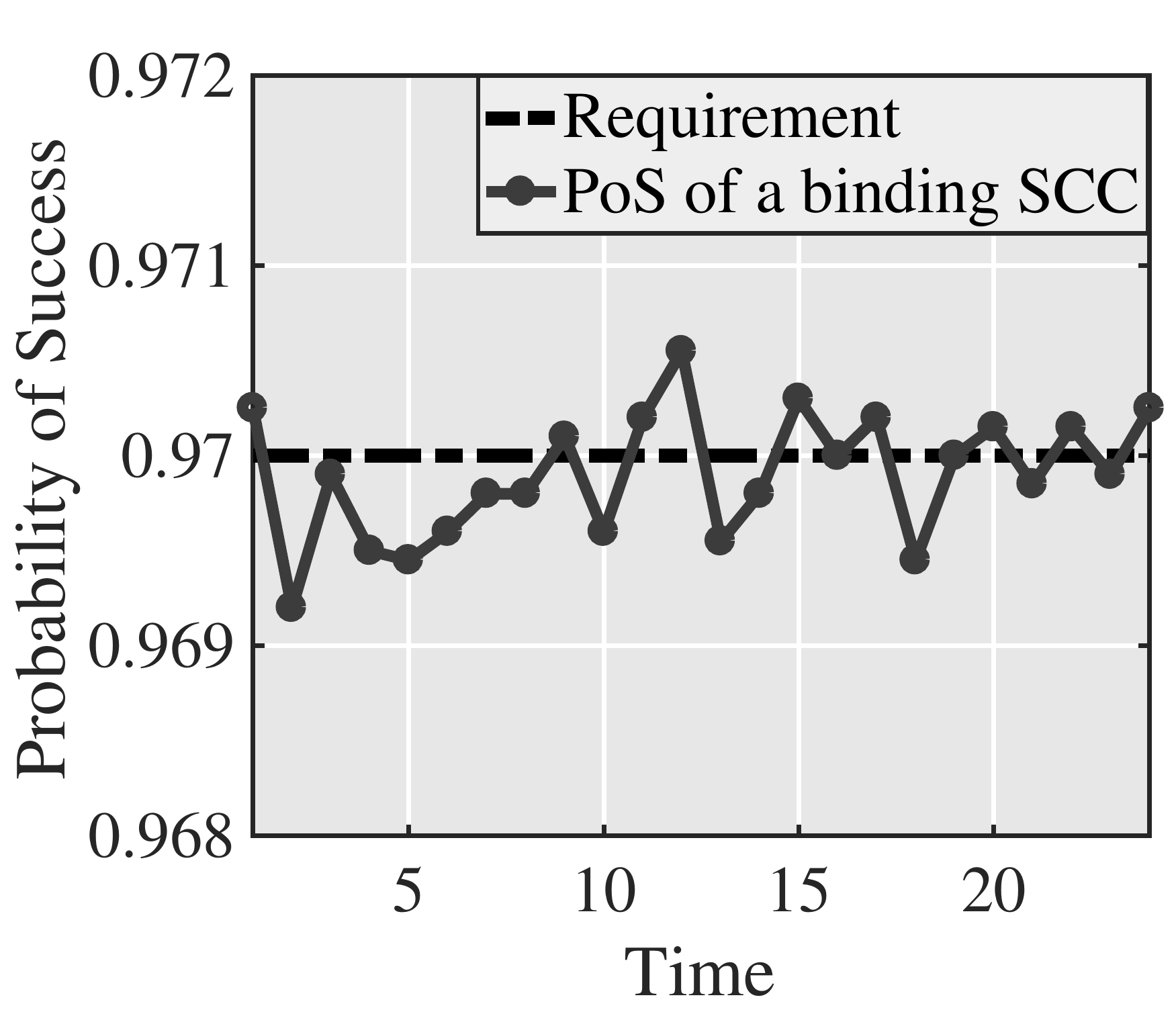} 
} 
\subfigure[]{ 
\includegraphics[width=1.61in]{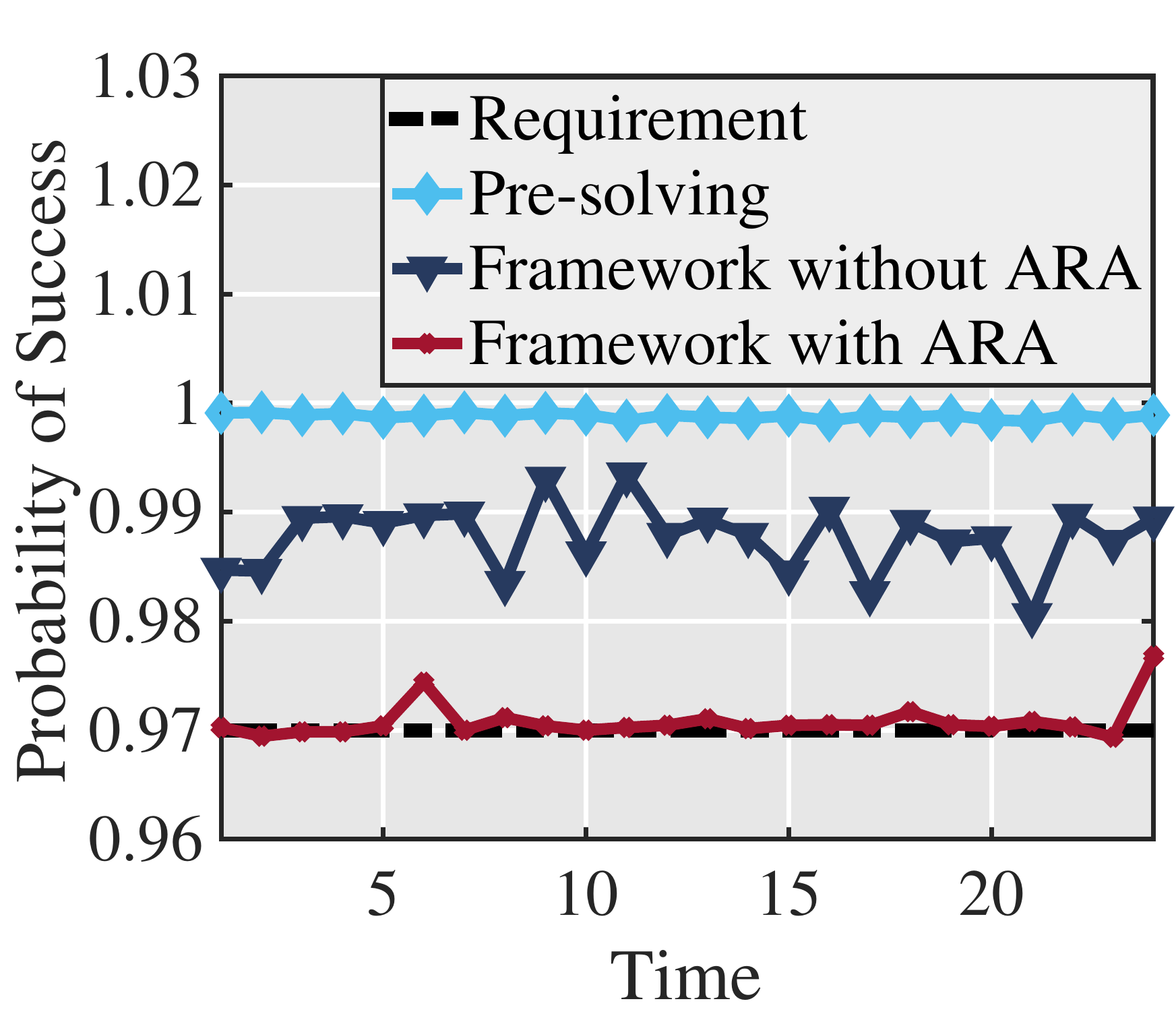} 
} 
\subfigure[]{ 
\includegraphics[width=1.64in]{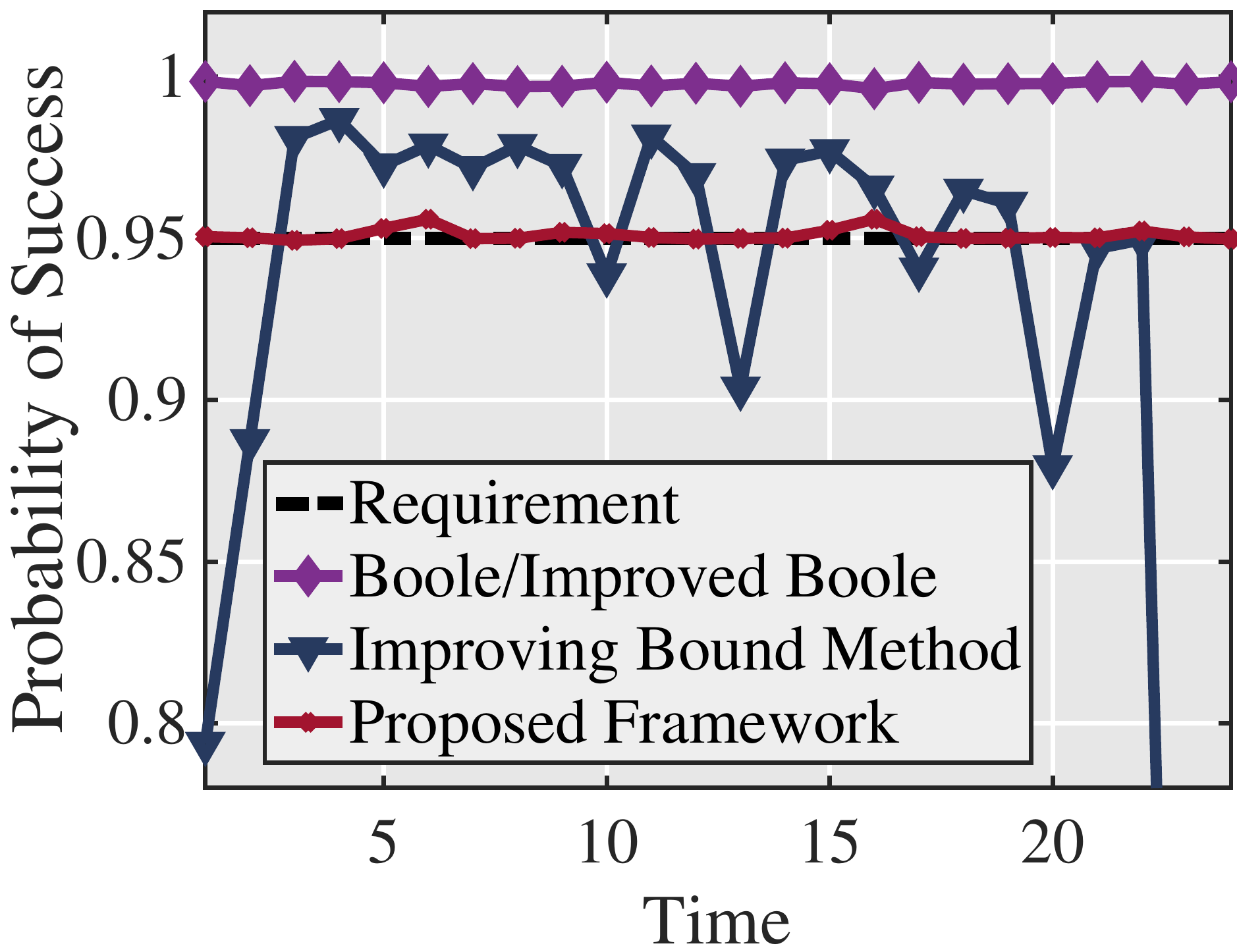} 
} 
\caption{(a) Evolution of the optimal value of the objective function (low, $\alpha=0.03$) (b) PoS of a binding SCC (low, $\alpha=0.03$) (c) PoS using the proposed framework (low, $\alpha=0.03$, ARA: adaptive risk allocation) (d) PoS of the evaluated methods (high, $\alpha=0.05$)}
\label{Inner} 
\end{figure}

First, the evolution of the optimal objective function value using the proposed framework is given in Fig. \ref{Inner}(a). Clearly, the optimal value decreases sharply and quickly reaches a stable value. After 7 iterations totaling 82.6 seconds, the relative variation of the optimal value is less than $10^{-5}$. 

Then, Monte Carlo simulation is carried out to evaluate the \textit{Probability of Success} (PoS), which is the left-hand side of the JCC at each time instance and is enforced to be greater than or equal to $1-\alpha$. Ideally, the PoS without conservativeness should be equal to $1-\alpha$ if the JCC is binding. However, due to the sampling error in the Monte Carlo simulation, the ideal coincidence may be unobtainable. For example, the PoS of a binding single chance constraint is illustrated in Fig. \ref{Inner}(b). Theoretically, this PoS should be always equal to $1-\alpha$, yet the practical PoS randomly deviates from the required limit. Similar violations will also occur in the following, which are irrelevant to the proposed framework.

To verify the proposed framework, the corresponding PoSs are demonstrated in Fig. \ref{Inner}(c). Obviously, the PoS when only performing the \textit{pre-solving} step is overly conservative. After conducting the framework without the adaptive risk allocation, the PoS has been reduced significantly due to the iterative classification and estimation. However, there is still an obvious gap between the PoS and the allowed limit. Once the adaptive risk allocation strategy is adopted, the PoS decreases to essentially coincide with the allowed limit.

The proposed framework is further compared with existing decomposition-based methods, including the most frequently used Boole's inequality \cite{GROSSO2014504}, the improved Boole's inequality \cite{BAKER2017230}, and the improving bound method \cite{8662704}. The PoSs of these methods are given in Fig. \ref{Inner}(d). Clearly, the PoS of Boole's inequality is most conservative. Besides, since the estimation of the joint violating probability of all the events is zero, the PoS of the improved Boole's inequality is equal to the result of Boole's inequality \cite{BAKER2017230}. In most cases, the PoS of the improving bound method is less conservative compared to Boole's/improved Boole's inequality. Nevertheless, in some time steps, the PoS of the improving bound method fails to meet the probability requirement. As mentioned earlier, the improving bound method ignores the interdependence among the classification, estimation, and solution. Thus, the reasons for the non-compliance are twofold. First, $E$ may be overestimated in these time instances. Second, the final solution is incompatible with the classification. In fact, the constraint of the reverse power flow on line 13 is classified as non-binding and removed from the joint CC-OPF for these time steps, but this flow does violate limits given the consequent solution, resulting in the PoS not meeting the requirement. Overall, the proposed framework shows superior performance compared to all the other evaluated methods, as the PoS of the framework almost coincides with the given limit, leaving nearly no room for improvement while still fulfilling the constraint.

The optimal generation cost of the evaluated methods are listed in Table \ref{fval}, except for the cost of the improving bound method because it intermittently fails to meet the requirement. Needless to say, the optimal cost when ignoring the JCC is the lowest, while the cost of Boole's/improved Boole's inequality is the highest. The proposed framework is a compromise between the two resulting in considerably lower cost than the conservative solution. 
\vspace{-0.5cm}
\begin{table}[h]
\setlength{\abovecaptionskip}{0pt}
	\renewcommand{\arraystretch}{1.4}
	\caption{Optimal Generation Cost of Evaluated Methods (high, $\alpha=0.05$)}
	\label{fval} 
	\centering
	\footnotesize
	\setlength{\tabcolsep}{1.6mm}{
	\begin{tabular}{l c c c}
	\hline
	\bfseries Method & Without JCC & Boole/Improved Boole  & Proposed Framework \\
	\hline
	\bfseries Cost  & 30863.4 & 33123.7 & 32273.1  \\
	\hline
	\end{tabular}}
\end{table}

\vspace{-0.5cm}
\section{Conclusion}
For joint CC-OPFs, this letter proposes a non-parametric iterative framework to decompose JCCs. The framework also contains an original adaptive risk allocation strategy. Results show that both the iterative structure and adaptive risk allocation can reduce conservativeness, while the latter contributes more significantly. Overall, the framework nearly eliminates the conservativeness, thereby cutting back the generation cost considerably. Although this framework can converge in practice and offer satisfactory solutions given different system parameters, theoretical proof of its convergence may reveal more insights, which is the future work of this letter.

\vspace{-0.2cm}
\bibliographystyle{IEEEtran}
\bibliography{IEEEabrv,paper}
\end{document}